\documentclass{article_saj}
\usepackage{graphicx,saj,multicol,subeqnarray}
\usepackage{natbib}
\usepackage{float}
\usepackage{xcolor}
\usepackage{widetext}
\usepackage{url}
\usepackage{bm}
\usepackage{tikz} 
\usepackage{pifont} 
\usepackage{amsfonts}
\usepackage{amssymb}
\usepackage{amsmath,upgreek}
\usepackage{titlesec}
\usepackage{ulem}
\titlelabel{\thetitle.\quad}
\definecolor{xlinkcolor}{cmyk}{1,0.6,0,0}
\usepackage[bookmarks=false,         
     pdfnewwindow=true,      
     colorlinks=true,    
     linkcolor=xlinkcolor,     
     citecolor=xlinkcolor,     
     filecolor=xlinkcolor,  
     urlcolor=xlinkcolor,      
final=true
]{hyperref}

\def\papertype{\ \hfill\ Invited Review}

\def\udc{52}
\begin{document}
\parindent=.5cm
\baselineskip=3.8truemm
\columnsep=.5truecm
\newenvironment{lefteqnarray}{\arraycolsep=0pt\begin{eqnarray}}
{\end{eqnarray}\protect\aftergroup\ignorespaces}
\newenvironment{lefteqnarray*}{\arraycolsep=0pt\begin{eqnarray*}}
{\end{eqnarray*}\protect\aftergroup\ignorespaces}
\newenvironment{leftsubeqnarray}{\arraycolsep=0pt\begin{subeqnarray}}
{\end{subeqnarray}\protect\aftergroup\ignorespaces}
%


\markboth{\eightrm THE CRITICAL MASS RATIO FOR W UMA-TYPE SYSTEMS}
{\eightrm B. ARBUTINA {\lowercase{\eightit{and}}} S. WADHWA}

\begin{strip}

{\ }

\vskip-1cm

\publ

\type

{\ }


\title{THE CRITICAL MASS RATIO FOR W~UMA-TYPE CONTACT BINARY SYSTEMS}


\authors{Bojan Arbutina$^{1}$ and Surjit Wadhwa$^2$}

\vskip3mm


\address{$^1$Department of Astronomy, Faculty of Mathematics,
University of Belgrade\break Studentski trg 16, 11000 Belgrade,
Serbia}


\Email{arbo@astro.edu.rs}

\address{$^2$School of Science, Western Sydney University, Locked Bag 1797, Penrith, NSW 2751, Australia}

\Email{19899347@student.westernsydney.edu.au}


\dates{March 31, 2024}{March 31, 2024}


\summary{Contact binaries are close binary systems in which both components fill their inner Roche lobes so that the stars are in direct contact and in potential mass and energy exchange. The most common such systems of low-mass are the so-called W UMa-type. In the last few years, there is a growing interest of the astronomical community in stellar mergers, primarily due to the detection of gravitational waves (mergers of black holes and neutron stars), but also because of an alternative model for type Ia supernovae (merger of two white dwarfs), which are again particularly important in cosmology where they played an important role in the discovery of dark energy and the accelerated expansion of the Universe. In that sense, contact systems of W UMa-type with extremely low mass ratio are especially interesting because there are indications that in their case, too, stars can merge and possible form fast-rotating stars such as FC Com stars and the blue-stragglers, and  (luminous) red novae such as V1309 Sco. Namely, previous theoretical research has shown that in the cases when the orbital angular momentum of the system is only about three times larger than the rotational angular momentum of the primary, a tidal Darwin's instability occurs, the components can no longer remain in synchronous rotation, orbit continue to shrink fast and they finally merge into a single star. The above stability condition for contact systems can be linked to some critical mass ratio below which we expect a system to be unstable. We give an overview of this condition and show how it can be used to identify potential mergers. Finally, we discuss a number of known extreme mass ratio binaries from the literature and prospect for future research on this topic.}



\end{strip}

\tenrm


\section{INTRODUCTION}

   Contact systems represent close binary systems in which both components fill their inner Roche lobes so that the stars are in direct contact and in a potential exchange of mass and energy. Although there are massive OB, even O-O contact binaries \citep{2021A&A...651A..96A, 2022A&A...666A..18A}, the most common such low-mass systems are the so-called stars of W Ursae Majoris (W UMa) type systems with components of late spectral classes, which have a common convective envelope and approximately the same effective temperatures \citep{a5}.

   In the last few years, special attention of astronomers has been attracted by stellar mergers, primarily due to the detection of gravitational waves (mergers of black holes and neutron stars) \citep{2016PhRvL.116f1102A, 2017PhRvL.119p1101A}, but also as an alternative model for type Ia supernovae (merger of two white dwarfs), which are again particularly significant in cosmology where they played an important role in the discovery of dark energy and the accelerated expansion of the Universe \citep{1998AJ....116.1009R, 1999ApJ...517..565P}. It was initially believed that contact W UMa-type binary systems will dominate the Galactic gravitational wave background at low frequencies (detectable by future Laser Interferometer Space Antenna - LISA), however, it was realized later that it will, most probably, be completely dominated by detached and semidetached (AM CVn-type) white dwarf binary systems \citep[see][and references therein]{2014LRR....17....3P}. However, contact systems of type W UMa with a low mass ratio are still particularly interesting because there are indications that in their case, merger of stars can also occur and a possible formation of fast-rotating stars of type FK Com and the so-called "blues stragglers", and (luminous) red novae.

   Namely, earlier theoretical research has shown that in the case when the orbital angular momentum is only about three times greater than the rotational angular momentum of the primary component, tidal instability occurs \citep{d1}, the components can no longer remain in synchronous rotation, they rapidly spirally approach each other and finally merge into a single star. This stability condition for contact systems can be connected to the existence of some critical mass ratio below which we expect the system to be unstable.

   This paper is organized as follows: in Section 2 we give a brief overview of W UMa-type binaries, review and reanalyze the condition for their stability (Subsection 2.1), and, finally, compare the theoretical results with the observational data, with an aim to identify potential merger candidates (Subsection 2.2); in Section 3 we discuss prospects for future research and in Section 4 we give a conclusion.


\begin{figure*}[h!]
\centerline{\includegraphics[width=0.80\textwidth, bb=0 170 2379 2900,keepaspectratio=true]{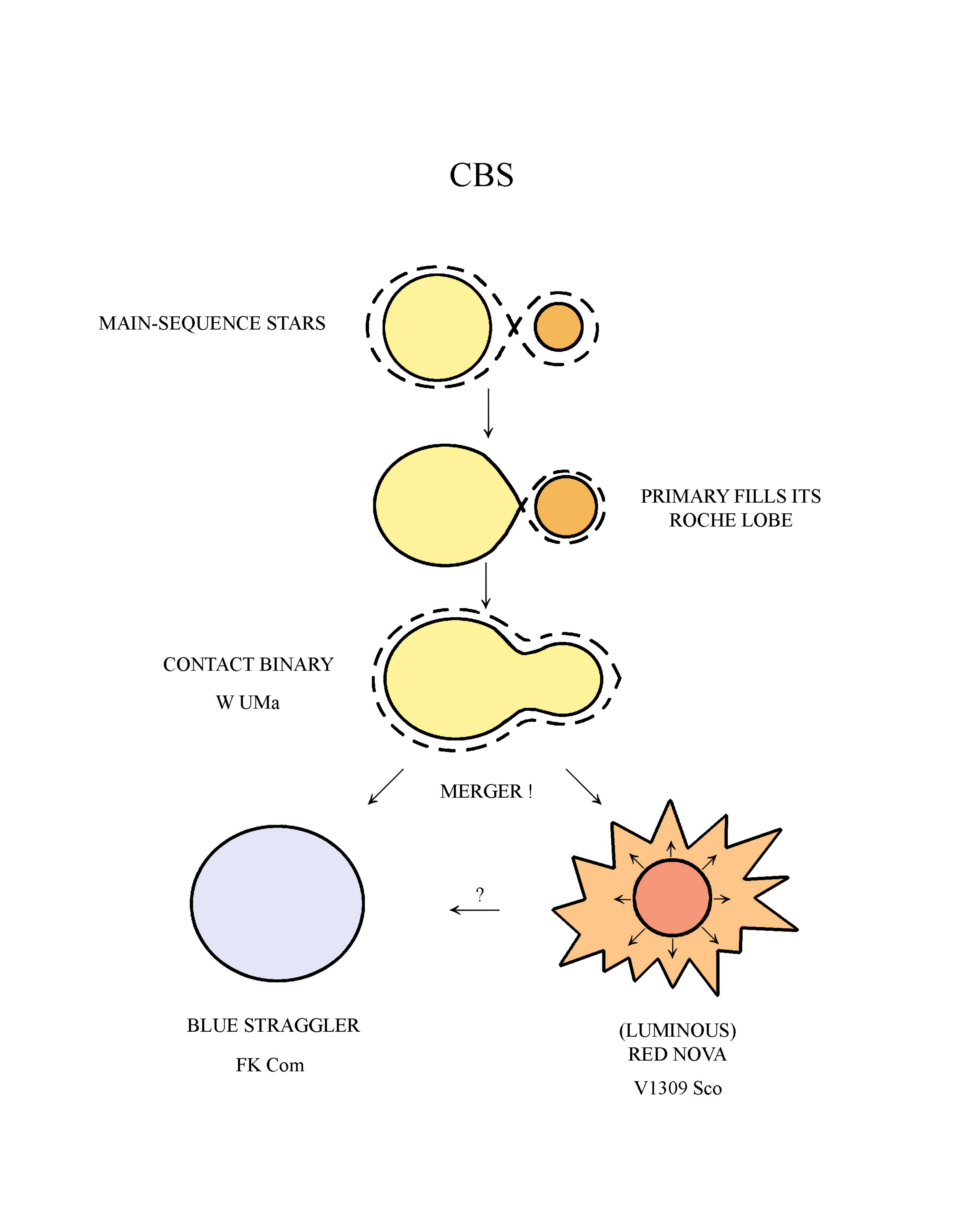}}
\caption{A possible evolution of W UMa-type close binary systems (CBS) cartoon.}
\label{Fig1}
\end{figure*}
\section{CONTACT BINARIES OF W UMA-TYPE}

W UMa-type binary systems were named after its prototype W UMa, an eclipsing binary of spectral type F5 V discovered by \citet{1903ApJ....17..201M}, with mass ratio $q = 0.508$, and primary mass $M_1 = 1.14\ M_\odot$ \citep{2021MNRAS.501.2897G}. W UMa binaries were defined as a class
in the reviews of \citet{1965VeBam..27...36B, 1970VA.....12..217B, 1977VA.....21..359B} and \citet{1985ibs..book...85R, 1985ibs..book..113R}. These are stars of spectral type late F--K, which have a common convective envelope and nearly equal effective temperatures, although the mass of the primary component is usually significantly greater than that of the secondary (typically twice i.e. $q =M_2/M_1 \sim 0.5$, but up to ten times or more).  Primary components in W UMa binaries seems to be normal main-sequence (MS) stars, while secondaries are oversized for their zero-age main-sequence (ZAMS) masses, and can be found left from the MS \citep{1993ASSL..177..111R, a5}. The mass transfer in contact binaries, whether from primary to secondary or vice versa, does not seem to be that large (on short timescales, at least), but the energy i.e. luminosity transfer needs to be substantial, in order to equalize effective temperatures of quite different stars, perhaps involving differential rotation \citep{a8, c1}.

There are two sub-types of W UMa binaries: A and W; the stars being classified as the former or the latter sub-type depending on whether the primary or the secondary is eclipsed during the primary (deeper) minimum, and consequently  whether the primary or the secondary has slightly higher temperature. We thus arrive at another peculiarity observed in W sub-type W UMa-type binaries, that the secondary, less massive star is hotter.
{In rare cases it seems, somewhat contradictory, that secondary is slightly hotter in A sub-type systems \citep{2007ASPC..370..279G,2016AcA....66..357A}. Spot presence and their specific locations may account for this A/W-subtype ambiguity \citep{2021AcA....71..123A}.}
There are also some other suggested sub-types. \citet{1979ApJ...231..502L} introduced the class of B-type systems which are systems in geometrical contact, but not in
thermal contact and therefore there are large surface temperature differences between the components.  H sub-type (high mass ratio) systems were introduced by \citet{2004A&A...426.1001C}. Low mass contact binaries of W UMa--type also exhibit a short period cut-off at $P \sim 0.15-0.22$ days whose origin is still under debate \citep{1992AJ....103..960R, 2006AcA....56..347S, 2020MNRAS.497.3493Z, 2022MNRAS.514.5528L, 2023MNRAS.523.1394Z, 2023ApJ...952..141P}.

 W UMa binaries are likely formed from detached systems of low mass, with orbital periods $P \lesssim$ 1 day, which lose angular momentum through some mechanism, probably magnetized stellar wind. Magnetized wind, and generally magnetic activity and presence of starspots i.e.  \citet[][]{1951PRCO....2...85O} effect, are also characteristic of W UMa systems \citep{a5}. As the orbit shrinks, primary will first touch its Roche lobe and transfer relatively small amount of mass to the MS secondary before stars get into contact (see Fig.~\ref{Fig1}). The system can stay in contact for an uncertain time interval in the so-called thermal-relaxation-oscillations (TRO) regime \citep{1976ApJ...205..217F, 1976ApJ...205..208L}. During TRO, system should actually oscillate between contact and marginally detached configuration, remaining in quasi-equilibrium. Mass exchange first assumes the flow from secondary to primary leading to the orbit widening and the loss of contact, but the primary would then expand, mass transfer from primary to secondary via Roche lobe overflow (RLOF) will lead to orbit shrinking, the contact is re-established and everything repeats from the beginning \citep{1993ASSL..177..111R, a5}. Alternatively, mass transfer via RLOF with mass ratio reversal may have happened prior to the first contact, implying that present secondaries are in an advanced evolutionary stage \citep{2006AcA....56..199S}.

 In any case, due to the angular momentum loss (AML), eventually the stars would merge, perhaps forming rapidly rotating object such as blue stragglers. In favor of this scenario is a number of W UMa-type binary systems among blue stragglers in open and globular clusters \citep{b2}. If a blue straggler is formed in stellar merger, is its formation preceded by red nova event such as V1309 Sco \citep[Fig. \ref{Fig1},][]{2019MNRAS.486.1220F}?
Alternatively (or predominantly) blue stragglers in old open clusters may form via mass-transfer from an asymptotic giant branch (AGB) or red giant branch (RGB) companion \citep{2019ApJ...881...47L, 2021ApJ...908..229L}. There is also a possibility for direct collisions in high stellar concentration cores of globular clusters, i.e. close encounters of single stars that will end in a coalescence. Of course, it is possible that some of these interesting objects, particularly those in young stellar environments, do not originate in merger or collisions, but in starbursts or delayed star formation \citep{1989AJ.....97..431E}.

\subsection{Stability criterion and the instability mass ratio}

We have seen that the long-term dynamical evolution of W UMa binaries is driven presumably by AML. In close binary systems,  tidal forces lead to synchronization and circularization of orbit. If the timescale for the synchronization is smaller that the AML timescale, the system will remain synchronized and orbit will shrink until, at some critical separation, the instability sets in -- the so-called secular, tidal or \citet{d1} instability. At this point rotational and orbital angular momentum become comparable  \citep[$J_\mathrm{orb} \approx 3 J_\mathrm{spin}$,][]{1980A&A....92..167H}. The system can no longer stay synchronized, and since angular momentum is still lost, the orbit will rapidly shrink until the final merger.

As we mentioned, the stability condition for contact systems can be linked to some minimum mass ratio below which we expect a system to be unstable \citep{a1, a2, b1, bb2, bb3, 2010MNRAS.405.2485J,zz,Z2023}.
\citet{a1} showed that the minimum mass ratio will depend on the dimensionless gyration radius of the primary, defined through the moment of inertia of a star $I = k_1^2 M_1 R_1^2$. In other words, the stellar structure determines the gyro-radius that enters into the calculation of the minimum mass ratio. For a fully radiative primary, taken to be $n=3$--polytrope, Rasio's expression $ {a}/{R_1} = k_1 \sqrt{{3(1+q)}/{q}}$, for a system in marginal contact ($R_1$ taken to be the mean radius of the inner Roche lobe for the primary and $k_1^2 =0.075$), gives $q_\mathrm{min} = 0.085$. Nevertheless, it was already known at the time about the extremely low mass ratio contact binary AW UMa with $q = 0.075$ \citep{1992AJ....104.1968R}. In order to place the AW UMa just at the stability boundary, it needed $k_1^2 \approx 0.06$, implying that its primary cannot have much of the convective envelope
and must be slightly evolved \citep{a1}. But, there were other extremely low mass ratio systems discovered, such as V857 Her with $q=0.065$) \citet{a10}.

On the theoretical ground, contribution of the rotational angular momentum of the secondary and inclusion of gyro-radius $k_2$ was considered by \citet{a2}. \citet{b1}, in addition to this, taken into account the fact that radii $R_2$ and $R_1$ in contact binaries are correlated, and through this correlation there was an additional dependence of angular momentum on binary separation.
Taking into account the structure of the primary components (deformation of the primary due to rotation and companion presence), \citet{bb2} included in the calculation nonzero quadruple moment i.e. apsidal motion constant, which slightly improved the $q_\mathrm{min}$ value but the problem remained. However, the analysis of \citet{bb2} was performed for idealized fully radiative primary, i.e. $n=3$--polytrope, while it was know that evolved MS stars with $M_1 \gtrsim \ 1\ M_\odot$ can have lower gyro-radius \citep{a2, 2010MNRAS.405.2485J}. \citet{2010MNRAS.405.2485J} reanalyzed the minimum mass ratio by considering the structure of MS stars using Eggleton’s stellar evolution code \citep{1971MNRAS.151..351E, 1972MNRAS.156..361E, 1973A&A....23..325E, 2002ApJ...575..461E}, emphasizing the importance primary's gyro-radius $k_1$.  By analyzing statistically empirical relations for deep, low mass ratio contact binaries \citet{2015AJ....150...69Y} concluded that $q_\mathrm{min}$ could be as low as 0.044. \citet{zz} constructed $k_1 - M_1$ relation for ZAMS stars based on \citet{2009A&A...494..209L} calculations for rotationally and tidally distorted components in close binaries.

In the following we shall repeat the stability analysis of \citet{b1} and \citet{zz} -- assume a certain dependence of filling factor on stellar volume radii, fit the gyro-radius--mass dependence for ZAMS primaries and derive an improved theoretical stability condition. We start from the total angular momentum of a binary
\begin{equation}
J_{\mathrm{tot}} = J_{\mathrm{orb}} + J_{\mathrm{spin}} = J_{\mathrm{orb}} + J_1 + J_2,
\end{equation}
where $J_1$ and $J_2$ are spin angular momenta of the components.
The orbital angular momentum of a binary can be written as
\begin{equation}
J_{\mathrm{orb}} =  \mu a^2 \Omega = \frac{q\ \sqrt{G M^3
a}}{(1+q)^2},
\end{equation}
where the reduced mass is $\mu = M_1 M_2 /M$, the total mass is $M=M_1+M_2$, the mass ratio $q = M_2/M_1 < 1$, and $M_1$
and $M_2$ are masses of the primary and secondary component,
respectively. $\Omega = \sqrt{GM/a^3}$ is the (Keplerian) orbital angular velocity, while $a$ is binary separation.
Assuming synchronization, the spin angular momentum of a binary
 is
\begin{equation}
J_{\mathrm{spin}} =  k_1^2 M_1 R_1^2 \Omega + k_2^2 M_2 R_2^2
\Omega,
\end{equation}
where $R_1$ and $R_2$ are taken to be the volume radii \citep[see][]{a6}, and $k_1$, $k_2$ are gyro-radii.

The overcontact degree for a contact binary is defined as
\begin{equation}
f =  \frac{\Phi _\mathrm{eff} - \Phi _{\mathrm{IL}}}{\Phi _{\mathrm{OL}} - \Phi
_{\mathrm{IL}}} .
\label{eq4}
\end{equation}
Value $f=0$ corresponds to marginal contact (components reaching L1 point), while $f=1$ corresponds to full overcontact configuration (the secondary reaching L2 point).
\citet{a8} adopted
logarithmic, \citet{b1} adopted linear dependence of $ \Phi
_{\mathrm{eff}}$ on $R$, while \citet{bb2} assumed $\Phi _{\mathrm{eff}} \propto 1/R$. We adopt the last dependence, which seems reasonable, because if
 $\Phi \propto 1/{r}$ the effective potential should be well
represented by a similar dependence on volume ("effective") radius.
Nevertheless, the exact dependence is not that important in the narrow range $f=0-1$ where
all approximations give similar accuracy.

\begin{figure}
   \centering
   \includegraphics[width=\columnwidth,bb=50 50 750 650,keepaspectratio=true]{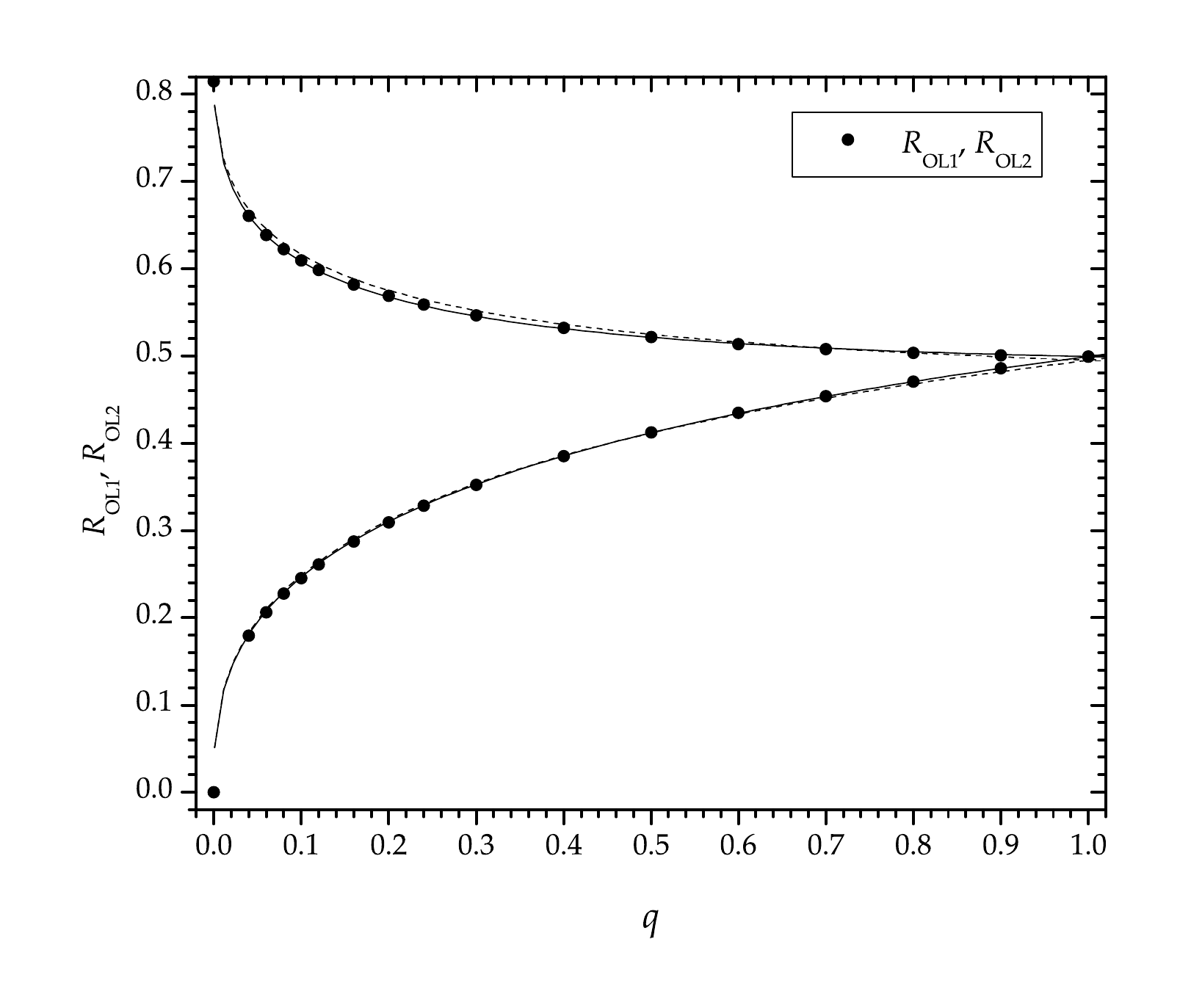}
   \caption{Mean radii for outer Roche lobes $R_{\mathrm{OL}1}$ and $R_{\mathrm{OL}2}$. Filled circles are numerical data from \citet{a6} tables, dashed curves are approximations given by \citet{a8}, while solid curves represent our approximations from Eq. (\ref{eq7}). }
              \label{Fig2}%
    \end{figure}

Thereby, from Eq. (\ref{eq4}) we have
\begin{equation}
f \approx \frac{1/R - 1/R _{\mathrm{IL}}}{1/R
_{\mathrm{OL}} - 1/R _{\mathrm{IL}}},
\end{equation}
where volume
radii for the inner Roche lobes, touching at L1, are approximately \citet{a7},
\begin{equation}
  \frac{R_{\mathrm{IL}i}}{a} = \Bigg\{ \begin{array}{ll}
 { \frac{0.49q^{-2/3}}{0.6q^{-2/3} + \ln (1+ q^{-1/3})}, } &  i=1 \\
 { \frac{0.49q^{2/3}}{0.6q^{2/3} + \ln (1+ q^{1/3})}, }  &  i=2,
 \end{array}
\end{equation}
while for the volume radii of the outer Roche lobes we suggest
\begin{equation}
\frac{R_{\mathrm{OL}1}}{a} = \frac{0.49q^{-2/3}\big(\cosh (1.15 q^{2/5})\big)^{1/2}}{0.6q^{-2/3} + \ln (1+ q^{-1/3})},
\label{eq7}
\end{equation}
\begin{equation}
\frac{R_{\mathrm{OL}2}}{a} = \frac{0.49q^{2/3}\big(1-q^{1/4}\tanh ^2 (1.15 q^{1/5})\big)^{-1/4}}{0.6q^{2/3} + \ln (1+ q^{1/3})}.\nonumber
\end{equation}
The latter are defined as the radii of the spheres, each being of
the same volume as the volume of the respective figure obtained by
cutting the equipotential surface passing through the L2 point by
a plane through the L1 point which is perpendicular to the line of
centers. Formulae in Eq. (\ref{eq7}) are slightly better approximations than those given by \citet{a8}, accurate to less than 1 per cent when compared to the \citet{a6} tables (Fig. \ref{Fig2}).

As the component's surfaces in the contact system are at the same
potential (the same $f$), the stellar radii are correlated and by combining the above equations one
obtains
\begin{equation}
R_2 = \frac{R_{\mathrm{IL}2} R_{\mathrm{OL}2}}{ f{R_{\mathrm{IL}2}} +(1-f) R_{\mathrm{OL}2}} ,
\label{eq8}
\end{equation}
where
\begin{equation}
f= \frac{1/R_1 - 1/R_{\mathrm{IL}1}}{1/R_{\mathrm{OL}1}
- 1/R_{\mathrm{IL}1}}.
\label{eq9}
\end{equation}

From the instability condition $\frac {\mathrm{d}
J_{\mathrm{tot}} }{\mathrm{d} (a/R_1)}$ = 0 one  finds the equation for the
critical separation
\begin{equation}
\bigg(\frac{a_{\mathrm{\scriptscriptstyle inst}}}{R_1}\bigg)^2 = \frac{1+q}{q} \bigg[ 3 k_1^2  - q  k_2^2 \bigg(\frac{R_2}{R_1}\bigg)^2  \bigg( 1 - 4 \frac{R_2}{R_1} S \bigg)\bigg] ,
\label{eq10}
\end{equation}
where
\begin{equation}
S(q) = \frac{1/R_{\mathrm{OL}2} - 1/R_{\mathrm{IL}2}}{1/R_{\mathrm{OL}1}
- 1/R_{\mathrm{IL}1}}.
\end{equation}
In a situation when the secondary (i.e. its angular momentum) has been neglected ($k_2=0$), the instability equation is reduced to
\begin{equation}
 \frac{a_{\mathrm{\scriptscriptstyle inst}}}{R_1} = k_1 \sqrt{\frac{3(1+q)}{q}},
\end{equation}
which is the result found in \citet{a1}, that we already quoted.

\begin{figure}
   \centering
   \includegraphics[width=\columnwidth,bb=50 50 750 650,keepaspectratio=true]{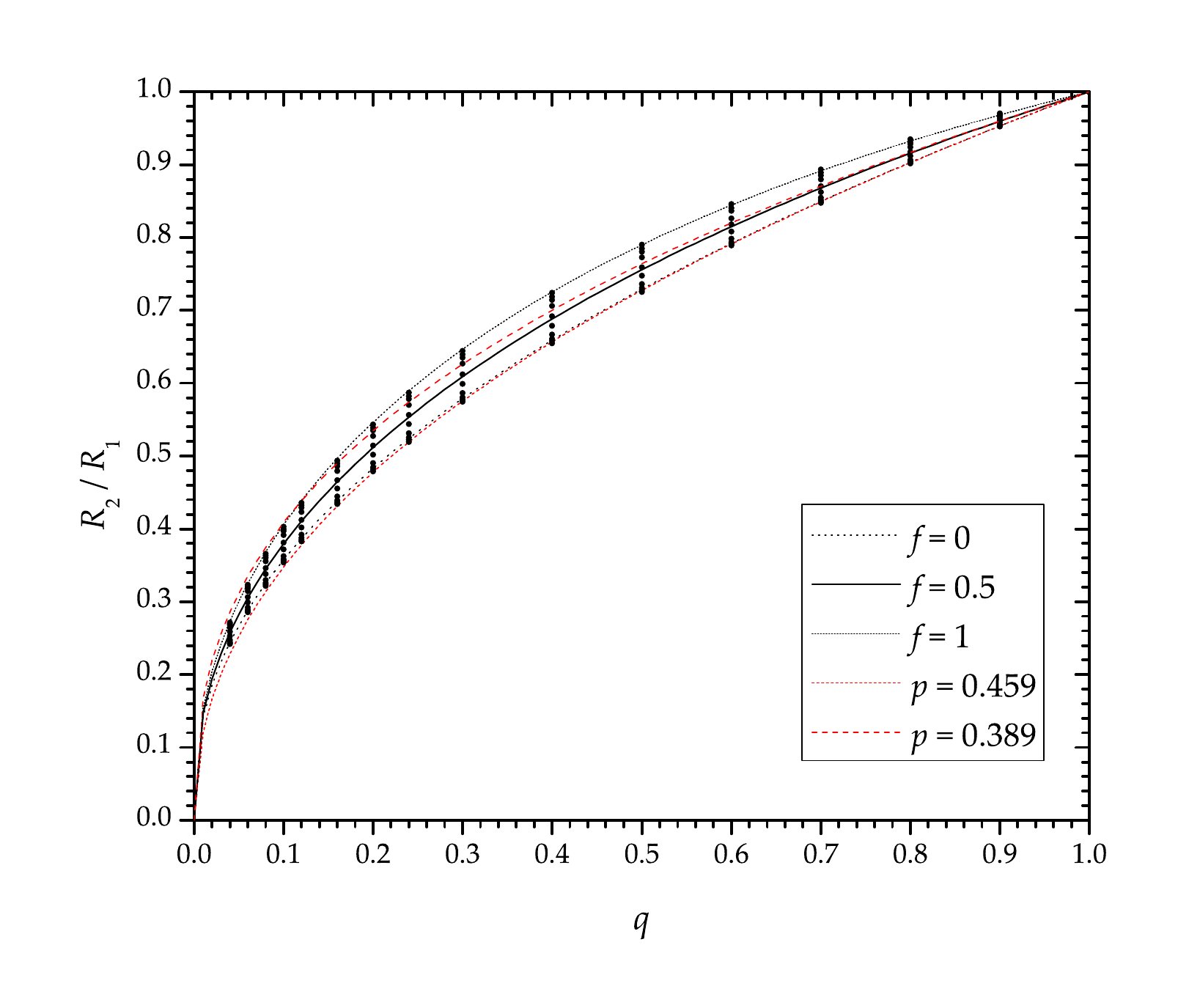}
   \caption{The numerical data for the ratio $R_2/R_1$ from \citet{a8} and our approximations. The dotted curve represents a power-law fits $R_2/R_1 \approx q^p$ to the observational data \citep[$p=0.459$,][]{1941ApJ....93..133K, 2004A&A...426.1001C} \citep[$p=0.389$,][]{2022PASP..134f4201P}. }
              \label{Fig3}%
    \end{figure}

By transforming Eq. (\ref{eq9}) to obtain equation equivalent to Eq. (\ref{eq8}) but for the primary
\begin{equation}
R_1 = \frac{R_{\mathrm{IL}1} R_{\mathrm{OL}1}}{ f{R_{\mathrm{IL}1}} +(1-f) R_{\mathrm{OL}1}} ,
\end{equation}
we obtain from the instability equation an algebraic equation for the instability mass ratio that depends on $k_1^2$, $k_2^2$ and $f$. Both above expressions for $R_1$ and $R_2$ are accurate to less than 1 per cent when compared to the data from \citet{a6}. The radii ratio in Eq.~(\ref{eq10}) is explicitly
\begin{eqnarray}
\frac{R_2}{R_1}
 &=& \frac{f \big(\cosh (\frac{23}{20} q^{2/5})\big)^{-1/2}+1-f}{f\big(1-q^{1/4}\tanh ^2 (\frac{23}{20} q^{1/5})\big)^{1/4} +1-f}\cdot \nonumber \\
  &\cdot& \frac{1 + \frac{5}{3} q^{2/3}\ln (1+ q^{-1/3})}{1 + \frac{5}{3} q^{-2/3}\ln (1+ q^{1/3})}
\end{eqnarray}
To lesser accuracy this ratio for a fixed filling factor (or for all $f$) could be represented by a power-law $R_2/R_1 = q^p$ (see Fig.~\ref{Fig3}).

\begin{figure}
   \centering
   \includegraphics[width=\columnwidth,bb=50 50 750 650,keepaspectratio=true]{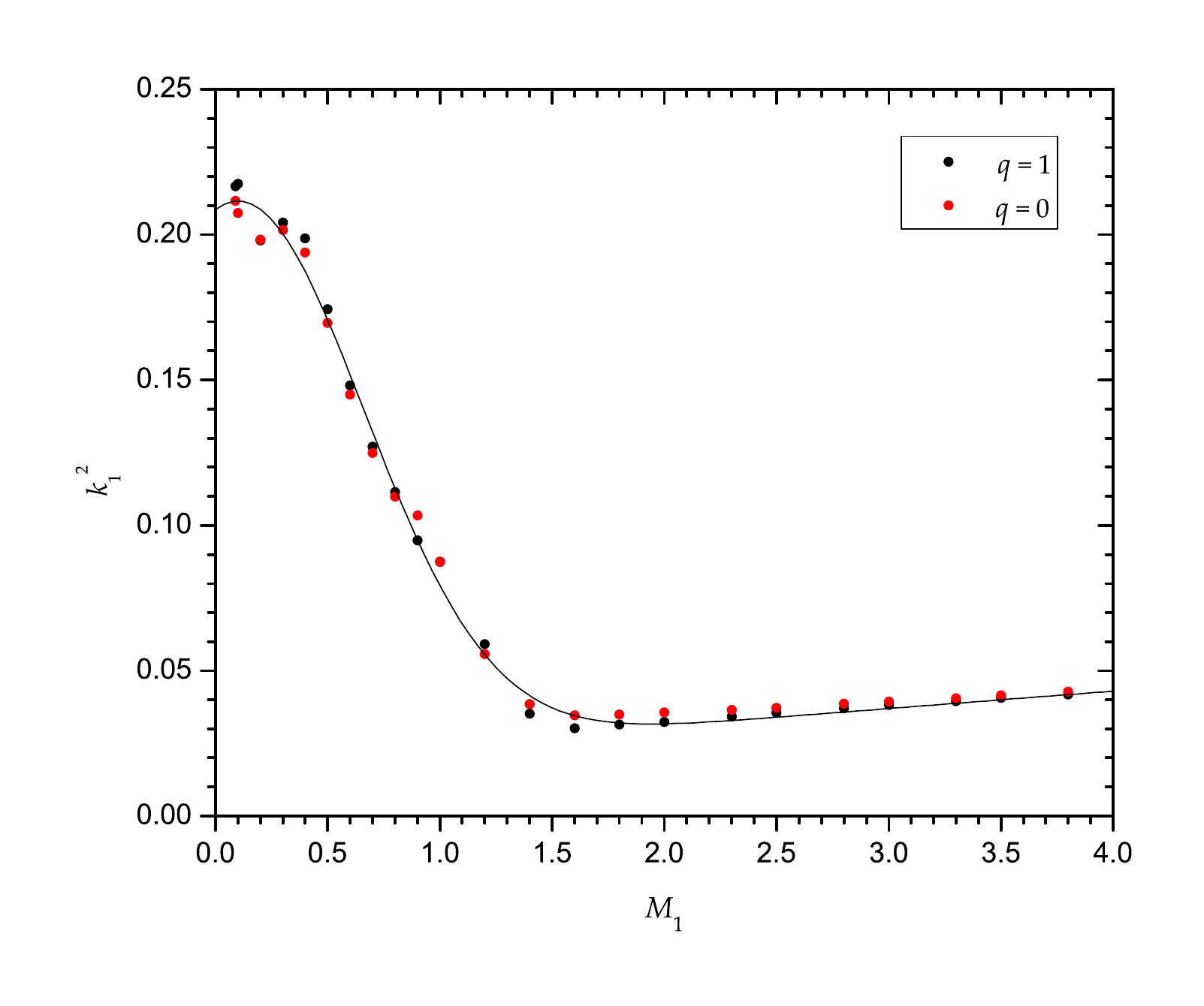}
   \caption{The data \citep{2009A&A...494..209L} and $k_1^2 = k_1^2 (M_1)$ relation fit. }
              \label{Fig4}%
    \end{figure}

Although it is clear from papers by \citet{2010MNRAS.405.2485J} and \citet{zz} that gyration radius $k_1$ depends on $M_1$ and have minimum at $\sim  1.5 M_\odot$ for ZAMS models  \citep[e.g.][]{2009A&A...494..209L}, we shall show this explicitly by fitting $k_1^2 = k_1^2 (M_1)$ relation (Fig. \ref{Fig4}).  Since \citet{2009A&A...494..209L} calculations for binary model assumed $q=1$, and we are dealing with low-mass ratio systems, we fitted both, the binary models data that include rotational and tidal effects and isolated star models ($q=0$, $\Omega=0$). We assumed a Gaussian + linear dependence of $k_1^2$:
\begin{equation}
 k_1 ^2 = C e^{-((M_1-m)/s)^2} + a M_1 +b  ,
\end{equation}
with best fit parameters given in Table \ref{table1}  ($M_1$ is in Solar mass units).

By using last equation and assuming a fully convective secondary ($n$=1.5--polytrope) with $k_2^2 = 0.205$, one can find the minimum mass ratio
\begin{equation}
q_\mathrm{min} = 0.042-0.044,
\end{equation}
for the filling factor $f=0-1$.
One must bear in mind that this is a global minimum, and $q_\mathrm{inst}$ that is of practical use is different for each particular binary \citep{2023PASP..135i4201W}.

\begin{figure*}[h!]
   \centering
   \includegraphics[width=0.80\textwidth,bb=50 50 750 643,keepaspectratio=true]{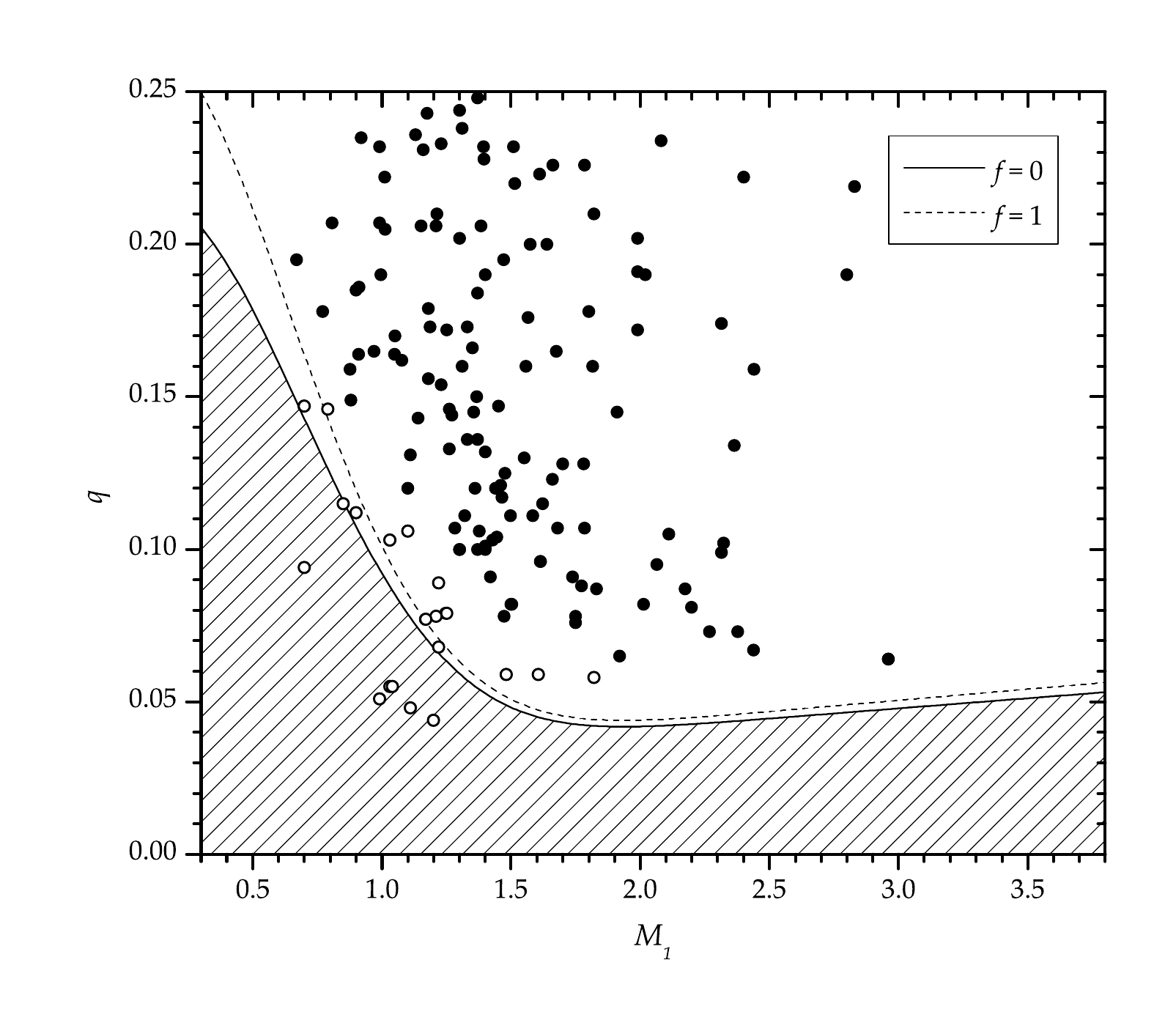}
   \caption{The instability mass ratio versus primary mass. The data are from \citet{2021ApJS..254...10L} (filled circles) and Table \ref{table2} (open circles),}
              \label{Fig5}%
\end{figure*}

The instability mass ratio versus primary mass for $k_2^2 = 0.205$ and $f=0-1$ is shown in Fig. \ref{Fig5}. The data are from \citet{2021ApJS..254...10L}.
Extreme and low mass ratio contact binaries that we consider as possible merger candidates are listed in Table \ref{table2} and included in Fig. \ref{Fig5} as well. Systems listed by \citet{2021ApJ...922..122L} that we did not include in the table since they are probably not merger candidates, based on our analysis (being relatively massive), are: V857 Her, M4 V53, V870 Ara, KR Com, FP Boo, KIC 11097678, XX Sex and AW Crv. Similar situation is probably with AW UMa ($q=0.076$) \citep{2016MNRAS.457..836E}, VESPA V22 ($q=0.079$) \citep{2022NewA...9701862P}, GSC 02265-01456 ($q=0.087$) \citep{2015JApA...36..399G},  NW Aps  ($q=0.10$) and AL Lep ($q=0.12$) \citep{2005Ap&SS.300..329W}. {Recently, \citet{2022MNRAS.512.1244C,  2023MNRAS.519.5760L, 2023PASP..135i4201W, 2023PASP..135g4202W} found a number of low mass ratio W UMa-type systems, none of them, however, satisfy our instability criterion, although some are close.} Even the majority of systems that we included in Table \ref{table2} seems to be stable, for now. Only nine fulfill the criterion for instability, having $q<q_\mathrm{inst}$. We discuss them and some other interesting systems in more detail in the following subsection.

%

   \begin{table}
      \caption[]{Parameters in the gyro-radius--primary's mass relation $k_1^2 (M_1) = C e^{-((M_1-m)/s)^2} + a M_1 +b $.}
         \label{table1}
         $$
         \begin{array}{p{0.3\linewidth}l}
            \hline
            \noalign{\smallskip}
            Parameter     &   \\
            \noalign{\smallskip}
            \hline
            \noalign{\smallskip}
            {\it C} & 0.192\pm 0.006   \\
            {\it m} & 0.09\pm 0.03            \\
            {\it s} & 0.81\pm 0.04  \\
            {\it a} &  0.006\pm 0.002             \\
            {\it b} &  0.019\pm 0.005            \\
            \noalign{\smallskip}
            \hline
         \end{array}
         $$
   \end{table}
%

%


\begin{table*}[t]
\caption{Extreme and low mass ratio contact binaries -- potential merger candidates. Instability mass ratio is given only for systems with $q<q_\mathrm{inst}$. }
\parbox{\textwidth}{
\vskip.25cm
\centerline{\begin{tabular}{@{\extracolsep{1.0mm}}lcccccccccccc@{}}
  \hline

        Name      & \multicolumn{3}{c}{} & Mass & & & & \multicolumn{2}{c}{Temperatures} &  & \ & {Ref.} \\
              && $q$ && $M_1$ [$M_\odot$] && $f$ & $i$ [$^\circ$] & {\footnotesize $T_1$ [K]} & {\footnotesize $T_2$ [K]} &  & $q_\mathrm{inst}$ &   \\
 \hline
 V1187 Her       && 0.044  && 1.20$^\mathrm{b}$  && 0.84 &66.7& 6250 & 6682 &  & 0.072 & (1)\\
 TYC 4002-2628-1 && 0.048  && 1.11  && 0.35 &69.7& 6032 & 6151 &  & 0.096 & (2)\\
 WISE J185503.7         && 0.051  && 0.99 && 0.16 &70.7& 5747 & 5827 &  & 0.095 &  (3)\\
 WISE J141530.7         && 0.055  && 1.04 && 0.64 &77.4& 5890 & 5966 &  & 0.091 &  (4)\\
 VSX J082700.8         && 0.055  && 1.03$^\mathrm{b}$ && 0.19 &68.7& 5870 & 5728 &  & 0.089 &  (5)\\
 IP Lyn$^\mathrm{a}$            && 0.058 && 1.82 && 0.22 &76.8& 6680 & 6180 & & -- &  (6)\\
 KIC 4244929    && 0.059  && 1.48  && 0.81 &70.6& 6857 & 6867 &  & -- & (7)\\
 KIC 9151972    && 0.059 && 1.61  && 0.76 &70.1& 6040 & 5982 &  & -- & (7)\\
 ASAS J083241+2332.4 && 0.068  && 1.22 && 0.69 &82.7& 6300 & 6672 &  & 0.069 &  (8)\\
 NSVS 2569022        && 0.077 && 1.17 && 0.01 &76.3& 6100 & 6100 &  & -- &  (9)\\
 ZZ PsA        && 0.078  && 1.21 && 0.97 &72.2& 6514 & 6703 &  & -- &  (10)\\
 SX Crv  && 0.079$^\mathrm{c}$ && 1.25 && 0.27 &61.2& 6340 & 6160 &  & -- &  (11)\\
 1SWASP J132829  && 0.089  && 1.22$^\mathrm{b}$ && 0.70 &81.5& 6300 & 6319 &  & -- &  (5)\\
 V1309 Sco$^\mathrm{d}$  && 0.094 && 0.7?$^\mathrm{e}$ && 0.89 &73.4& 4500 & 4354 &  & 0.161&  (12)\\
 ASAS J165139+2255.7  && 0.103 && 1.03 && 0.60 &78.1& 5370 & 5394 &  & --&  (13)\\
 ASAS J082243+1927.0  && 0.106 && 1.10 && 0.78 &76.6& 5960 & 6078 &  & --&  (14)\\
V1222 Tau  && 0.112 && 0.90 && 0.54 &80.2& 5425 & 5600 &  & 0.113&  (15)\\
 GSC 02800-01387  && 0.115 && 0.85 && 0.63 &74.6& 5684  & 5846 &  & 0.124&  (16)\\
 NSVS 1917038  && 0.146 && 0.79$^\mathrm{b}$ && 0.04 &73.8& 4869  & 5074 &  & --&  (17)\\
 NSVS 4316778  && 0.147 && 0.70 && 0.00 &75.6& 5960  & 6100 &  & --&  (18)\\
   \hline
\end{tabular}}}
\vskip 2mm
{\small
 $^\mathrm{a}$No--spots solution. $^\mathrm{b}$Main-sequence mass based on temperature from 2022 updated version of tables by \citet{2012ApJ...746..154P} and \citet{2013ApJS..208....9P}.$^*$  $^\mathrm{c}$Spectroscopic mass ratio. $^\mathrm{d}$Parameters before merger.  $^\mathrm{e}$Assigned mass for a fictitious main-sequence doppelganger.\\
References: (1) \citet{2019PASP..131e4203C}, (2) \citet{2022MNRAS.517.1928G}, (3) \citet{2023MNRAS.521...51G},  (4) \citet{2023PASP..135d4201G},  (5) \citet{2021ApJ...922..122L}, (6) \citet{2023RAA....23h5013Y},  (7) \citet{2016PASA...33...43S}, (8) \citet{2016AJ....151...69S}, (9) \citet{2018RAA....18..129K}, (10) \citet{zz}, (11) \citet{2004AcA....54..299Z},  (12) \citet{2016RAA....16...68Z}, (13) \citet{2018JAVSO..46....3A},   (14) \citet{2015MNRAS.446..510K}, (15) \citet{2015PASJ...67...74L},  (16) \citet{2022NewA...9701862P}, (17) \citet{2020MNRAS.497.3381G}, (18) \citet{2020NewA...7701352K}.\\
$^*$\href{http://www.pas.rochester.edu/~emamajek/EEM\_dwarf\_UBVIJHK\_colors\_Teff.txt}{http://www.pas.rochester.edu/\~{}emamajek/EEM\_dwarf\_UBVIJHK\_colors\_Teff.txt}}

\label{table2}
\end{table*}


\subsection{Interesting systems}

\subsubsection{FK Com}

HD 117555 was recognized to be rapidly rotating G-type star by \citet{1948PASP...60..382M}. \citet{1966IBVS..172....1C} showed this star to be (micro)variable with small brightness variation of $\sim 0.1$ magnitude and period of $P\approx 2.4$ days, after which it was designated FK Comae Berenices (FK Com). FK Com was soon suggested to be the prototype of a new class of variables, including UZ Lib and V1794 Cygni \citep{1981ApJ...247L.131B, 1981IAUS...93..177B, 1981A&A...104..260R}. It is a giant star (G2 III -- G7 III) with unusually fast rotation, with $v \sin i \approx 160$ km/s, large cool spots and chromospheric activity, and long term variability \citep{2007A&A...467..229P}. FK Com shows variable radio, X-rays, UV and H$\alpha$ emission  \citep{2005A&A...434..221K}.

In some characteristics (enhanced chromospheric, transition region and coronal emission) FK Com--type stars are similar to RS CVn--type (RS Canum Venaticorum) systems \citep{1988MNRAS.232..361M}. Nevertheless, FK Com displays a lack of an observable radial velocity variation due to binarity  and broad H$\alpha$ emission line with a strongly variable profile. Its binary nature is not completely excluded.  \citet{1982ApJ...260..735W} suggested that the giant may be accreting mass from a small unseen companion, but since the attempts to reveal the companion were unsuccessful, it would have to be a indeed very small, low-mass star \citep{1984ApJ...283..200M}. The main arguments for the binary model may be the long-term stability of the brightness variations, since dark starspots appearing always at the same locations i.e. longitudes in the single star model is not what one would expect \citep{1991AJ....101.2199R}. However, if FK Com was a low--mass ratio system it may be that the stars already merged i.e. that this interesting rapidly rotating giant is actually the result of merger of a W UMa-type contact binary \citep[][and references therein]{2016ApJS..223....5A}.

\subsubsection{AW UMa}

Paczyncki's star AW UMa (BD $+30^\circ$2163) was discovered in 1964 as a W UMa--type eclipsing binary with a period of about $P \approx 0.44$ days \citep{1964AJ.....69..124P}.
This was the first extremely low mass ratio contact binary with $q=0.075$ \citep{1992AJ....104.1968R} and for decades the record holder. However, \citet{b3} found higher mass ratio $q \sim 0.1$ and suggested that AW UMa  may not be a contact binary after all. In their new model AW UMa was a detached system, but with equatorial disk (belt) encompassing both components. This view was further supported by high-resolution spectroscopic observations by \citet{2015AJ....149...49R}, who suggested that AW UMa is a very tight, semi-detached binary in which there is a mass transfer from the more massive to the less massive component. \citet{2016MNRAS.457..836E} criticised the latter model, reanalyzed the system and provided an alternative solution, among others, in which AW UMa is a contact binary with $q = 0.076$, having polar spots. The last author concluded that a better approach to explain line profiles would be to consider differential rotation of both components \citep{2016MNRAS.457..836E}. The author, however, did not provide any direct evidence for the existence of spots and did not discuss the distinct possibility of the spotted solution being non-unique and potential existence of other solutions with significantly different geometric parameters \citep{1993A&A...277..515M, 1999NewA....4..365E}.

AW UMa shows steady period decrease $\dot{P} = - 0.145 \times 10^{-7}$  \citep{2008ApJ...672..575W}, but not actual signs of instability. From primary temperature $T_1 = 7410$ K, we estimate its mass $M_1 \approx 1.75\ M_\odot$ \citep{2012ApJ...746..154P, 2013ApJS..208....9P}. \citet{b3} find $M_1 \sim 1.5\ M_\odot$ but for lower temperature $T_1 = 6980$ K. In any case,
higher mass of the primary, in comparison the the Solar value, suggest that the system is not unstable, despite low mass ratio (see Fig. \ref{Fig5}).

\subsubsection{V1309 Sco}

V1309 Scorpii (V1309 Sco) was discovered as nova in 2008 by \citet{2008IAUC.8972....1N}. Soon it became clear that this was not a typical classical nova, but the so-called red nova \citep{2010A&A...516A.108M}, and was later characterised as Rosetta stone of contact binary mergers \citep{2011A&A...528A.114T, 2016A&A...592A.134T}.

Optical Gravitational Lensing Experiment (OGLE) observations exist for the star in the period 2001--2008, when the outburst happened \citep{2003AcA....53..291U}. \citet{2011A&A...528A.114T} find V1309 Sco progenitor to be
K1-3 III giant, in a system with initial period of about 1.44 days with exponential decay
\begin{equation}
P = P_0 \exp (\tau/(t-t_0)),
\end{equation}
where $t$ is time in Julian Dates (JD), $t_0 = 2455233.5$ and $\tau = 15.29$ and $P_0 =  1.4456$ days.
It may be that V1309 Sco was a contact binary of W UMa type, but as we stated earlier these systems generally have primaries that can be regarded as MS stars, and orbital periods which are typically less than a day, V1309 Sco then perhaps being at the long period cut-off \citep{1998AJ....115.1135R}?

Taking into account observed characteristics above, \citet{2011A&A...531A..18S} concluded that V1309 Sco was different from W UMa-type stars -- primary being a giant that recently filled its Roche lobe, and that the contact phase was very short, ending in merger.
\citet{2014ApJ...786...39N} modelled the system as contact binary having a sub-giant primary with $M_1 \approx 1.52 \ M_\odot$ and mass ratio $q = 0.105$. \citet{2011A&A...528A.114T} assumed lower mass $\sim 1\ M_\odot$. Since our analysis is for MS stars, to see where V1309 Sco would be on our $q-M_1$ plot (Fig. \ref{Fig5}), we assign to primary $M_1 \approx 0.7 \ M_\odot$ which is the MS mass corresponding to K-type star with $T_1 = 4500$ K. Binaries with less massive primary should not reach RLOF, and consequently contact configuration, within the age of the Universe \citep{2006AcA....56..347S}. Mass ratio $q=0.094$ \citep{2016RAA....16...68Z} for the assigned  primary mass would definitely put V1309 Sco in the instability region. For an evolved primary, to push the system at the edge of stability $q_\mathrm{inst}\approx q=0.094$ one would need gyro-radius $k_1^2 \sim 0.073 -0.081$ (see Fig. \ref{Fig6}).

\subsubsection{V857 Her}
V857 Herculis (V857 Her) is an extreme mass ratio contact binary with period of $P\approx 0.38$ days. \citet{2005AJ....130.1206Q} report the photometric analysis of V857 Her and derive a mass ratio of $q=0.065$ and high fillout of 83.8\%. They find a weak evidence that the orbital period may show a continuous increase at a rate of $\dot{P} = 2.9 \times 10^{-7} $ days/yr.
The authors did not provide an estimate of the mass of the primary. Using various published empiric relations such as the estimate of the absolute magnitude of the primary from the secondary eclipse \citep{zz} and mass-period relation \citep{2015AJ....150...69Y} we estimate the mass of the primary to be $1.3 - 1.5~M_\odot$. Adopting a mean value of $M_1 = 1.4~M_\odot$ we would consider V857 Her to be stable. Thus, the situation with this system may be similar to AW UMa case -- extremely low mass ratio system, but with relatively massive primary. In addition, it is possible that the light curve of V857 Her may be influenced by the presence of a hot sub-dwarf \citep{2009AJ....137.3655P}.

\subsubsection{V1187 Her}
V1187~Herculis (V1187~Her) was discovered by Robotic Optical Transient Search Experiment I (ROTSE I) and designated as ROTSE-1 J162919.83+353959.2 \citep{2000AJ....119.1901A}. It was classified as EW variable with amplitude of about 0.2 magnitudes and period $P\approx 0.31$ days.
With $q=0.044$ V1187 Her currently holds the record for the most extreme low mass ratio contact binary system \citep{2019PASP..131e4203C}. The system exhibits a period change at rate  $\dot{P} = -1.5 \times 10^{-7}$ days/year.

Although the original authors do not provide an estimate of the mass of the primary, spectroscopic classification and observational evidence would suggest the mass of the primary to be in the order of $M_1 = 1.1 -1.2~M_\odot$. The system would undoubtedly be classified as unstable at this estimate ($q_\mathrm{inst} = 0.072$ for $M_1 = 1.2$ and $f=0.84$). There is some evidence, however, suggesting that the system is contaminated by a significant third light with the most recent estimates of the mass ratio of the system may be as high as 0.16 \citep{2022AAS...24020505C, 2024AAS...24342302C}.

\subsubsection{TYC 4002-2628-1}

TYC 4002-2628-1 (CzeV710, WISE J230927.8 +545123) was discovered as EW variable with a period $P\approx 0.37$ days by \citet{Pintr}
\citep[see also][]{2017OEJV..185....1S}. Photometric observation and light-curve analysis by \citet{2022MNRAS.517.1928G} find TYC 4002-2628-1 to be an extreme low mass ratio system with $q=0.048$. The ephemeris
shows a secular period increase of $\dot{P} = 1.62 \times 10^{-5}$ days/year. This high long-term period increase was suggested by authors to be a consequence of mass transfer from secondary to primary star. For the assumed primary mass $M_1 = 1.11~ M_\odot$ and fillout $f=0.35$, the instability mass ratio is $q_\mathrm{inst} = 0.096$. For the above system parameters, TYC 4002-2628-1 is thus significantly below the stability limit and should be unstable.

\subsubsection{WISE J185503.7+592234}

WISE J185503.7+592234 (ASASSN-V J185503.69 +611804.1, ZTF J185503.69+611804.2) was discovered by Zwicky Transient Facility (ZTF) \citep{2020ApJS..249...18C}. \citet{2023MNRAS.521...51G} recently observed and analyzed the system, finding it to be a particularly low mass ratio ($q\approx0.051$) contact binary
approaching merger. The period of the binary is $P\approx 0.28$ days with secular period decrease of $\dot{P} = -2.24 \times 10^{-7}$ days/year. The authors interpret this period change by mass transfer from primary to secondary, leading to even smaller mass ratio, deeper contact and eventual coalescence. For estimated primary mass $M_1 \approx 0.99~M_\odot$ and the fillout $f=0.16$, the mass ratio  is already significantly below the critical mass ratio $q_\mathrm{inst} = 0.095$ and the system should be unstable.

\subsubsection{WISE J141530.7+592234}

WISE J141530.7+592234 (ASASSN-V J141530.72 +592234.6, ZTF J141530.72+592234.3) was recognized as a contact binary with a low amplitude of $\sim$ 0.2 magnitudes and a short period $P\approx 0.34$ days
in Wide-field Infrared Survey Explorer (WISE) Catalog of Periodic Variable Stars \citep{2018ApJS..237...28C} and ZTF \citep{2020ApJS..249...18C}.
\citet{2023PASP..135d4201G} find system to be an extreme low mass ratio binary with $q\approx0.055$. The system displays significant light curve variations and \citet{1951PRCO....2...85O} effect reversal.
The period increase rate of $\dot{P} = 3.90 \times 10^{-7}$ days/year is tentatively explained by mass transfer from low mass secondary to the more massive primary component. For estimated primary mass $M_1 = 1.04~M_\odot$ and the filling factor $f=0.64$ for a hot spot solution, the critical mass ratio is $q_\mathrm{inst} = 0.091$, making the system clearly unstable.  This conclusion would not change for cool spot and no-spot solutions.

\subsubsection{VSX J082700.8+462850}

VSX J082700.8+462850 (VSX J082700) and 1SWASP J132829.37 +555246.1  are two extreme low mass ratio contact binaries analyzed by \citet{2021ApJ...922..122L}. While the latter has higher mass ratio and probably more massive primary, our analysis suggest that the former should be unstable.

VSX J082700 was first classified as EW variable by \citet{Srdoc}.
Photometric light curve solution of the system provided by \citet{2021ApJ...922..122L} suggest a mass ratio of $q = 0.055$. The system has a period $P\approx 0.28$ that shows decrease at the rate $\dot{P}= -9.52 \times 10^{-7}$ days/year. Based on photometric, color and empiric relationships the estimated mass of the primary range from 1.03 to $1.15~M_\odot$. For primary mass $M_1 = 1.03 \ M_\odot$, which is the MS mass corresponding to $T_1 = 5870$ K, and $f=0.19$, the instability mass ratio is $q_\mathrm{inst} = 0.089$. This places the system in the unstable category, similarly to V1187 Her. We find no evidence of blending from a nearby star, however, as it is the case with V1187 Her, this system has a very low inclination and as such contributing third light probably needs further investigation.

\subsubsection{ASAS J083241+2332.4}
{
ASAS J083241+2332.4 (NSVS 7399728, GSC 01941-02356) was observed with
Kilodegree Extremely Little Telescope (KELT) by \citet{2008AJ....135..907P}, classified as EB and designated KP301148. \citet{2016AJ....151...69S} photometrically observed and analyzed the system and found it to be an extreme low mass ratio contact binary. The system has a period $P\approx 0.31$ days with secular increase at the rate of $\dot{P} \sim 0.0765$ sec/year and sinusoidal modulation with a period of $\sim 8.25$ years, possible due to the presence of a third body. The latter authors find photometric mass ratio $q = 0.068$ for a solution with hot spot. The cool spot and no-spot solutions have slightly lower mass ratios, 0.065 and 0.067, respectively. Taking for the mass of the primary $M_1 = 1.22$ and filling factor $f=0.69$ for the hot spot solution, we find the critical mass ratio $q_\mathrm{inst}=0.069$. This makes the system unstable, but barely. Lower filling factor $f\approx 0$ and higher primary mass would make it stable.
}

\subsubsection{NSVS 2569022}

\citet{2006AJ....131..621G} classified NSVS
2569022 as a variable of EW type with period $P\approx 0.29$ days
and amplitude of about $\sim 0.2$ magnitudes.
\citet{2018RAA....18..129K} provide photometric solution for the light curve with estimated mass ratio of 0.077. It is difficult to estimate the mass of the primary component due to non availability of GAIA distance estimate and no high cadence photometry in V band. The reported mass of the primary, $1.17M_\odot$, based on the period-mass relationship, would place the system near the instability boundary, but stable ($q_\mathrm{inst}=0.071$). Although \citet{2024AAS...24342302C} do not provide an estimate of the mass ratio for the system, they state that the system contains a third light far more extreme than previously thought and does not have an extreme low mass ratio.

\subsubsection{ZZ PsA}

ZZ Piscis Austrinus (ZZ PsA) is a neglected bright southern contact binary recognized as a
variable in 1967 \citep{1967IBVS..195....1S}, rediscovered by \citet{1996IBVS.4322....1D}, and designated NSV 13890.
The light curve was analysed by \citet{2006Ap&SS.301..195W}, who found mass ratio  $q=0.080$ and  fillout of 90\%. The system has a period $P\approx 0.38$ days. No
period change was reported, owning to the lack of observations.
The system was recently analyzed by \citet{zz} reporting a mass ratio of 0.078 with higher fillout and estimated mass of the primary, based on the apparent magnitude of the secondary, as $M_1 = 1.21~M_\odot$.  For the above primary mass and filling factor $f=0.97$, the critical mass ratio for ZZ PsA is $q_\mathrm{inst}=0.072$. This makes the system stable, although it is close to the instability region in Fig.~\ref{Fig5}.

\subsubsection{V1222 Tau}

V1222 Tauri (V1222 Tau, GSC 00650-00769) is an ignored low mass ratio
contact binary found by \citet{2002IBVS.5234....1B}. The system has a period $P\approx 0.29$ days, with a possible secular increase rate of $\dot{P} = 8.19 \times 10^{-6}$ days/year \citep{2015PASJ...67...74L}.
\citet{2015PASJ...67...74L} provides both unspotted and spotted solution for the light curve. Adopting the spotted solution with $q=0.112$,  the reported mass of the primary $M_1 = 0.9~M_\odot$  and $f=0.54$, we find the critical mass ratio $q_\mathrm{inst}=0.113$, making system marginally unstable. For the unspotted solution the authors find even lower mass ratio ($q=0.104$) and higher fillout ($f=0.58$), which does not change much $q_\mathrm{inst}=0.114$ so the system remains unstable, for these parameter values.

\subsubsection{GSC 02800-01387}

GSC 02800-01387 (VSX J011323.6+374319) was discovered as EW variable by \citet{Miguel}.
It has a period $P\approx 0.3$ days.
\citet{2022NewA...9701862P} observed and provided light-curve fitting parameters for this and other three systems. They actually found VESPA V22 \citep{2017JAVSO..45...15Q} to be the most extreme mass ratio system of four targets, with $q = 0.079$, but with primary mass $M_1 = 1.99~M_\odot$ it falls in the stable region. The authors provide a spotted light curve solution for GSC 02800-01387 with an estimated mass ratio of $q=0.115$ and estimated mass of the primary $M_1=0.85~M_\odot$. The critical mass ratio for this primary mass and filling factor $f=0.63$ is $q_\mathrm{inst}=0.124$. Apart from concerns regarding the non-uniqueness of spotted solutions our analysis show this system to be unstable, for the current parameter values.

\subsubsection{NSVS 1917038}

NSVS 1917038 is discovered as a low mass ratio
binary ($q = 0.146$), with period  $P\approx 0.32$ days and an unusually shallow contact
degree of 4 per cent \citep{2020MNRAS.497.3381G}. We could not find much additional data for this system and assumed $M_1 = 0.79 \ M_\odot$ which is the MS mass corresponding to $T_1 = 4870$ K. For this primary mass and $f=0.04$, the critical mass ratio is $q_\mathrm{inst}=0.127$, which makes this system stable. This conclusion would change if $q$ and/or $M_1$ are slightly lower, and $f\lesssim 1$.

\subsubsection{NSVS 4316778}

NSVS 4316778 is an eclipsing binary with period $P\approx 0.26$ days
and amplitude of about $\sim 0.3$ magnitudes \citep{2004AJ....127.2436W}.
\citet{2020NewA...7701352K}  performed light curve analysis and found system to be basically in marginal contact, with photometric mass ratio $q=0.147$ and primary mass $M_1 \approx 0.7$.
Both components of NSVS 4316778 seems to be oversized, greatly overluminous and hotter
when compared to MS stars of the same masses. Although system show total eclipses, allowing for better parameters constrains, only a spotted solution is provided which makes the values somewhat unreliable. For the above parameters, assuming MS primary, the critical mass ratio for NSVS 4316778 is $q_\mathrm{inst}=0.143$, which makes it stable. However, greater fillout $f\lesssim 1$ would place the system below stability limit. The case of NSVS 4316778, as well as NSVS 1917038, demonstrates how relatively high mass ratio $q \sim 0.15$ in combination with low primary mass could make systems potentially unstable.

\begin{figure}[h!]
   \centering
   \includegraphics[width=\columnwidth,bb=50 50 750 650,keepaspectratio=true]{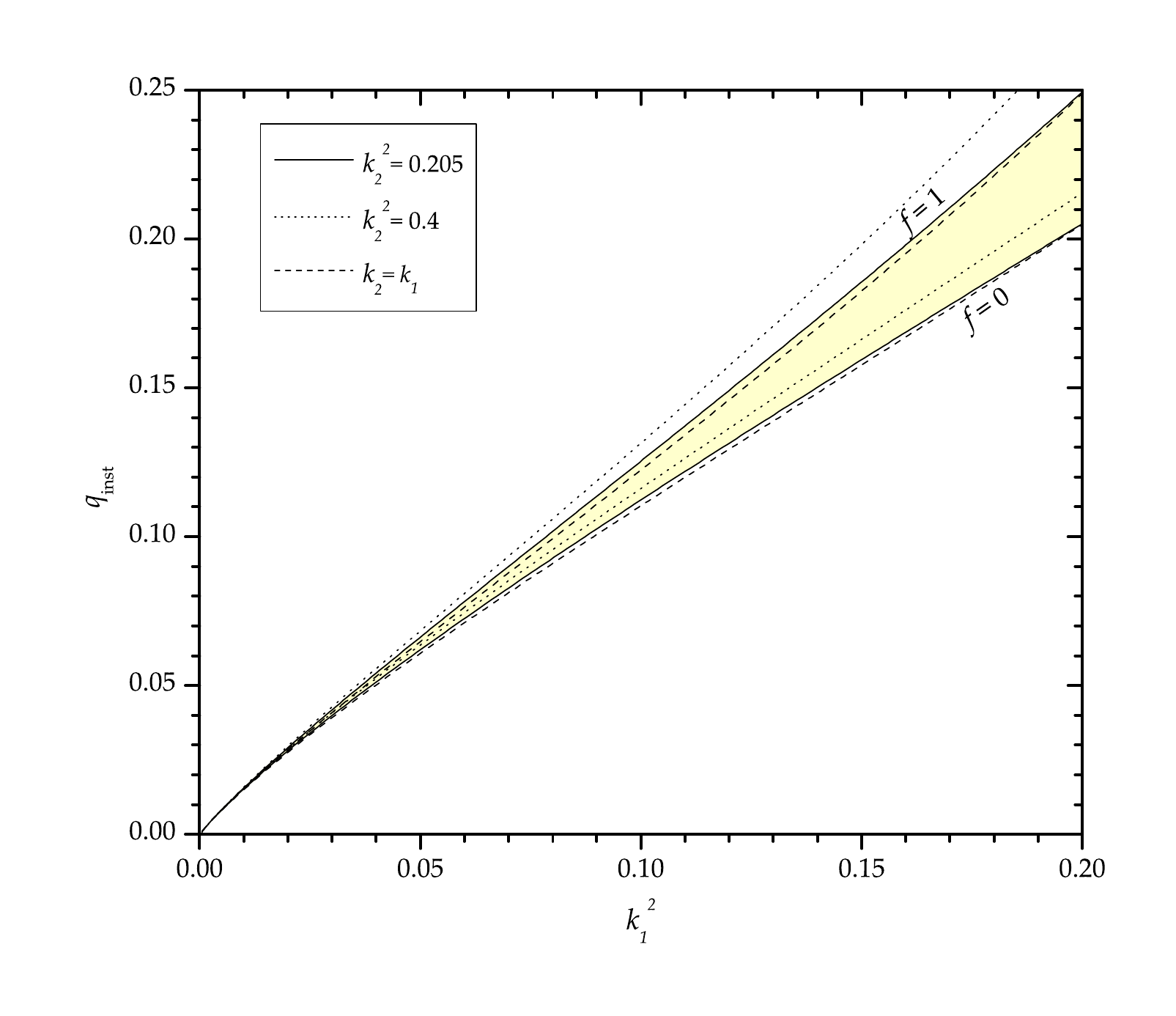}
   \caption{The instability mass ratio, according to Eq. (\ref{eq10}) versus primary's dimensionless gyration radius, with three cases assumed for the secondary: $n$=1.5--polytrope with $k_2^2=0.205$, homogeneous sphere with $k_2^2=\frac{2}{5}$ and $k_2 = k_1$. The lower curves correspond the filling factor $f=0$, while the upper curves correspond to $f=1$. }
              \label{Fig6}%
\end{figure}

\section{PROSPECTS FOR FUTURE\\ RESEARCH}

After a review of systems in the preceding section, an obvious question arises, why there are no more direct evidence of instability for binaries with $q < q_\mathrm{inst}$?
Concerning the observational data on low mass ratio W UMa-type binaries, one can hope that future research will provide more reliable system parameters, primary's mass $M_1$ and mass ratio $q$ in particular. To this task, high resolution spectroscopy would be very much needed. One should monitor closely the systems that are considered unstable and those at the border of instability region (in Fig.~\ref{Fig5}) e.g. search for rapid period change \citep{2024AJ....167...18H}, unusual brightness variation and other signs of instability. Comparison to observational characteristics of V1309 Sco and other possible Galactic red novae such as V4332 Sgr and V1148 Sgr \citep[][and references therein]{1999AJ....118.1034M, 2019A&A...630A..75P, 2022AJ....164...28B} can be extremely useful.

Theoretically derived critical mass ratio can be further improved by considering metallicity dependence. This has already been started by \citet{2024MNRAS.527....1W}. If age of a system could be estimated more reliably, evolutionary models for the primary i.e. its structure that determines gyro-radius $k_1$ could also come into play \citep{2010MNRAS.405.2485J}. Concerning the secondary's gyro-radius, it is also questionable whether $k_2$ for $n$=1.5--polytrope should be used, and could more adequate model for an oversized, tidally and rotationally distorted  star be constructed? \citet{2005MNRAS.360..272L} argued that the efficient energy transfer would decrease gyro-radius of the secondary and that the value of
$k_2$ would not be much different from $k_1$, despite a significant difference in stellar masses. {It is worth noting that for all the systems in Table \ref{table2} with mass ratio $q\leq 0.055$, the secondary's mass is less than the minimum mass for hydrogen fusion in isolated ZAMS stars, $M_2 <0.08\ M_\odot$.}

From a purely mathematical perspective, one could notice that in the derivation of criterion for tidal instability in Subsection 2.1, the total mass and mass ratio are treated as constants (we are basically applying the criterion for their current values), while this is surely physically unjustified -- more careful analysis should somehow try to account for mass transfer and/or mass loss. Observations suggest that mass ratio decreases during the course of time, perhaps not linearly but oscillatory, due to the TRO cycles, but W UMa binaries do pile up at small $q$ values \citep{2023A&A...672A.176P}. It is also possible that low-mass systems may reach full overcontact $f=1$, and that what drives instability then could be mass loss and AML through L2 \citep{1976ApJ...209..829W, 2011A&A...528A.114T, 2019MNRAS.489..891H}. This is a situation that should be further investigated, perhaps leading to a different criterion.

One interesting extension of the work on contact close binary systems stability would be a consideration of massive systems. Although theoretical investigations of massive contact binaries indicate
that these systems should tend to have mass ratio $q\simeq 1$, observational evidence show the opposite $q<1$ \citep[see][and references therein]{2022A&A...666A..18A}.
Stability analysis of massive binaries similar to the one for low-mass systems, however, bears some potential problems. One particular question is whether the standard Roche model is a valid description of such systems \citep{1972Ap&SS..19..351S, 1986Ap&SS.124....5D, 1995A&A...294..723D}?
But, even if we adopt this model, the primary may not be close to MS, that we generally assume for the low-mass systems, and the secondary will surely also be oversized but can no longer be treated as a low-mass fully convective star.

Nevertheless,
a number of objects, such as magnetic massive stars
\citep{2019Natur.574..211S}, Be stars \citep{2014ApJ...796...37S}, luminous
blue variables or similar stars such as $\eta$ Car \citep{2018MNRAS.480.1466S}, and peculiar Type-II supernovae like  SN1987A \citep{2017MNRAS.469.4649M} have all been suggested to result in massive binary mergers. An interesting example is V838 Mon. It appeared as atypical nova in 2002, later  characterized as luminous red nova (LRN). The erupting star was a cool extremely luminous supergiant \citep[in post-eruption phase designated as L-type][]{2003MNRAS.343.1054E}, with $L \gtrsim 10^6 L_\odot$ and radius reaching $R \gtrsim 10^3 R_\odot$ \citep{2005A&A...436.1009T}. The nova that made V838 Mon temporarily the brightest star in the Milky Way produced an iconic light echo
 \citep{2003Natur.422..405B} that helped to constrain the distance \citep{2004A&A...414..223T} and thus the absolute parameters, but the question of progenitor remained open. It is possible that
 nova V838 Mon was a merger in a triple system -- merging close binary consisting of B-type star + lower mass companion, in a binary orbit with another (survived) B-type star
 \citep[][and references therein]{2021A&A...655A..32K}. Similar events may be M31-RV \citep{1989ApJ...341L..51R} and  M31LRN~2015 \citep{2015ApJ...805L..18W, 2017MNRAS.470.2339L} in Andromeda galaxy (M31), and a small number of other extragalactic transients designated as LRN detected so far \citep{2019A&A...630A..75P, 2020MNRAS.492.3229H}.

\section{CONCLUSION}

Investigation of low mass ratio contact binaries of W UMa--type is a fruitful field of research, especially today when there is a growing interest of astronomical community in binary mergers.
As we have shown, under right conditions, a W UMa--type binary can reach critical separation, which can be related to the critical mass ratio below which we expect components to merge \citep{b1, bb2, bb3}. It is likely that some blues stragglers, FK Com-type stars and (luminous) red novae are produced in this way.

The critical mass ratio depends on primary's (and secondary's) gyro-radius (Fig. \ref{Fig6}). Assuming that the primary is a ZAMS star, we can relate its gyro-radius to the mass through $k_1^2 = k_1^2 (M_1)$ relation. This relation shows that there is a minimum of $k_1$ at about $ 1.5 M_\odot$ \citep{zz}, translating into minimum mass ratio
$$q_\mathrm{min} = 0.042-0.044,$$
depending on the filling factor ($f=0-1$), for roughly that same mass $ \sim 1.5 M_\odot$ (Fig.~(\ref{Fig5})).

One should keep in mind that this is a global minimum, and $q_\mathrm{inst}$ which decides whether the system is stable or unstable is different for each particular binary \citep{2023PASP..135i4201W} i.e. each primary, in this simplified analysis. Gyro-radius--stellar mass relation should also include metallicity dependence \citep{2024MNRAS.527....1W}. However, this relation is for ZAMS, and even if we neglect the question of the secondary, we could ask ourselves whether the primary is always close to ZAMS?  Gyration radius can be even lower for evolved MS stars \citep{2010MNRAS.405.2485J}. For example, for the Sun $k_\odot ^2 \approx 0.06$ \citep{a15}, while it is higher for 1 $M_\odot$ ZAMS star ($k_1^2 \approx 0.087$).
It is thus not impossible that there exist stable systems with mass ratio $q < q_\mathrm{min}$, but since most of the primaries in W UMa-type stars seem to be close to the MS, $q_\mathrm{min}$ should be a reasonably good estimate (as well as the values for $q_\mathrm{inst}$).

On the observational side, there is a number of past or ongoing searches for merging systems \citep{2017BlgAJ..26...26K, zz, 2021ApJ...922..122L, 2021MNRAS.502.2879G, 2022AJ....164..202L, 2022RAA....22j5009W, 2022JApA...43...94W, 2022JApA...43...42W, 2022MNRAS.512.1244C,  2022NewA...9701862P, 2023MNRAS.519.5760L, 2023PASP..135i4201W, 2023PASP..135g4202W}. We can hope that in the near future one of these searches will result in an identification of an unstable system, such as {V1309 Sco} \citep{2011A&A...528A.114T}, that will allow us to have another nature's live broadcast of a stellar merger event.

%
%
%


\acknowledgements{
    BA acknowledges the funding provided by the Science Fund of the Republic of Serbia through project \#7337 "Modeling Binary Systems That End in Stellar Mergers and Give Rise to Gravitational Waves" (MOBY), and by the Ministry of Science, Technological Development and Innovation of the Republic of Serbia through the contract \# 451-03-66/2024-03/200104.
}


\vskip.7cm
{\noindent\tenrm ORCiD\vskip.2cm}

\noindent Bojan Arbutina {\small \href{https://orcid.org/0000-0002-8036-4132}{https://orcid.org/0000-0002-8036-4132}}\\
\noindent Surjit Wadhwa {\small \href{https://orcid.org/0000-0002-7011-7541}{https://orcid.org/0000-0002-7011-7541}}
\vskip.7cm



\newcommand\eprint{in press }

\bibsep=0pt

\bibliographystyle{aa_url_saj}

{\small

\bibliography{wuma}
}

\clearpage

{\ }


\begin{strip}

{\ }






\vskip3mm







\vskip3mm







\baselineskip=3.8truemm

\begin{multicols}{2}

{
\rrm

{\ }

}

\end{multicols}

\end{strip}


\end{document}